\documentclass[lettersize,journal]{IEEEtran}
\usepackage{amsmath,amsfonts}
\usepackage{algorithmic}
\usepackage{algorithm}
\usepackage{array}
\usepackage[caption=false,font=normalsize,labelfont=sf,textfont=sf]{subfig}
\usepackage{textcomp}
\usepackage{stfloats}
\usepackage{url}
\usepackage{verbatim}
\usepackage{graphicx}
\usepackage{cite}
\usepackage{amsmath,amssymb,amsfonts}
\usepackage{algorithm, algorithmic, comment}
\usepackage{graphicx}
\usepackage{textcomp}
\usepackage{epsfig}
\usepackage{xcolor}
\usepackage{balance}
\newtheorem{theo}{Theorem}

\begin{document}

\title{Provable Privacy Guarantee for Individual Identities and Locations
in Large-Scale Contact Tracing}

\author{Tyler Nicewarner, Wei Jiang, Aniruddha Gokhale, Dan Lin,~\IEEEmembership{IEEE senior member}

}

\markboth{}%
{Shell \MakeLowercase{\textit{et al.}}: Provable Privacy Guarantee for Individual Identities and Locations
in Large-Scale Contact Tracing}


\maketitle

\begin{abstract}
The task of infectious disease contact tracing is crucial yet challenging, especially when meeting strict privacy requirements. Previous attempts in this area have had limitations in terms of applicable scenarios and efficiency. Our paper proposes a highly scalable, practical contact tracing system called PREVENT that can work with a variety of location collection methods to gain a comprehensive overview of a person's trajectory while ensuring the privacy of individuals being tracked, without revealing their plain text locations to any party, including servers. Our system is very efficient and can provide real-time query services for large-scale datasets with millions of locations. This is made possible by a newly designed secret-sharing based architecture that is tightly integrated into unique private space partitioning trees. Notably, our experimental results on both real and synthetic datasets demonstrate that our system introduces negligible performance overhead compared to traditional contact tracing methods. PREVENT could be  a game-changer in the fight against infectious diseases and set a new standard for privacy-preserving location tracking.
\end{abstract}


\section{Introduction}
	
	Infectious diseases have been a grave threat to public health for centuries. The COVID-19 pandemic has demonstrated the devastating impact such diseases can have on human lives and economies. In order to curb the spread of a highly contagious virus like COVID-19, it is essential to identify and quarantine individuals who may have been exposed to the virus through contact tracing. Various automated contact tracing methods have been proposed to reduce the burden on healthcare professionals. However, achieving full-spectrum identity and location privacy while collecting pervasive data and conducting big data analysis remains a major challenge. Unfortunately, existing industrial and academic works have failed to provide satisfactory solutions to this problem. Google and Apple's COVID-19 contact tracing apps, for instance, have been heavily criticized by security experts for their potential to enable advertisers to track users \cite{TradingPrivacy, tang2020privacypreserving}. Moreover, the apps have a critical limitation in that they are unable to identify individuals who have no direct contact with patients but have been exposed to the virus lingering in the air. Some virus like COVID-19 can remain airborne in confined spaces for up to three hours, which means individuals who visit a location shortly after a patient has left are still at risk of contracting the virus, even if they were never in close proximity with the patient. While recent work has attempted to address this issue by using QR code scanning to record visits to locations \cite{QRcode}, such approach is limited to places with QR codes. Undoubtedly, in order to obtain a comprehensive overview of an individual's trajectory, it may be necessary to utilize multiple methods of location collection in a coordinated manner. Nonetheless, to the best of our knowledge, there is no solitary contact tracing system that is capable of integrating location data obtained through diverse channels, while simultaneously ensuring privacy protection.


	
Despite the existence of several works that appear to address similar topics, such as location privacy protection \cite{MoveWithMe}, privacy-preserving trajectory querying \cite{PosTransform}, and privacy-preserving trajectory publishing \cite{TrafficInfo}, none of them are capable of effectively addressing the pressing privacy protection challenges involved in contact tracing. This is due to the fact that most existing location privacy-related works only require users to disclose either their real identities or approximate locations but not both simultaneously. For example, these works may enable a server to provide local weather forecasts to an anonymous user who discloses only their city but not their exact locations or analyze traffic flows on anonymized trajectories without revealing the owners of the trajectories. However, the unique requirements of virus tracking introduce new challenges for privacy protection since it necessitates querying both the real identity and precise location information while safeguarding both identity and location privacy. Without such protection, it becomes impossible to identify and notify individuals who may have been exposed to the virus.
	
In this  paper, we introduce PREVENT (Privacy pREserving Virus ENcountering Tracking), an innovative and  practical privacy preserving contact tracing system. PREVENT not only identifies people who have been in contact with a patient, but also  provides provable  privacy guarantees for both identities and locations. Our system enables a comprehensive recording of participant trajectories while upholding their privacy, and effectively addresses scalability issues, thereby bridging a significant gap in the current state of the art. PREVENT is composed of three main parties as illustrated in Figure \ref{fig:system}: (i) servers which host the system; (ii) subscribers such as health care providers, organizations, universities, companies, etc. which help verify patients' real identities (but not locations); (iii)  users who participate in contact tracing.  This system has the potential to revolutionize contact tracing by increasing efficiency and privacy protection of individuals. These advancements are expected to encourage more widespread adoption of contact tracing services.
The specific data flow in the PREVENT system is as follows. People who are willing to  contribute to the contact tracing install  the PREVENT mobile app. The PREVENT system will collect users' encrypted locations along with their periodically changing pseudo IDs from the mobile app through a variety of outdoor and indoor positioning systems that are available.  If a user  has been diagnosed and the user wishes to inform others who may be infected, the user can ask his/her  health care provider or employer who are subscribers of the service to   send a contact tracing request to the PREVENT system. The system  will conduct analysis directly on encrypted user data and  broadcast a list of pseudo  IDs of the people  who may have been exposed to virus to all the  subscribers. Each subscriber  maintains a list of real and pseudo identities of users who registered with it, and the subscriber will help inform the users in the risk list.  During the whole process, both users' identities and trajectories are always anonymous to servers; location information about users including the patients is never disclosed to subscribers (e.g., their employers or health care providers); users only receive simple notifications about potential virus exposure but not any information about when/where they may have been in contact with which patients.  In summary, our work makes the following unique contributions:
	
	\begin{figure}[!t]
		\centering
		\epsfig{file=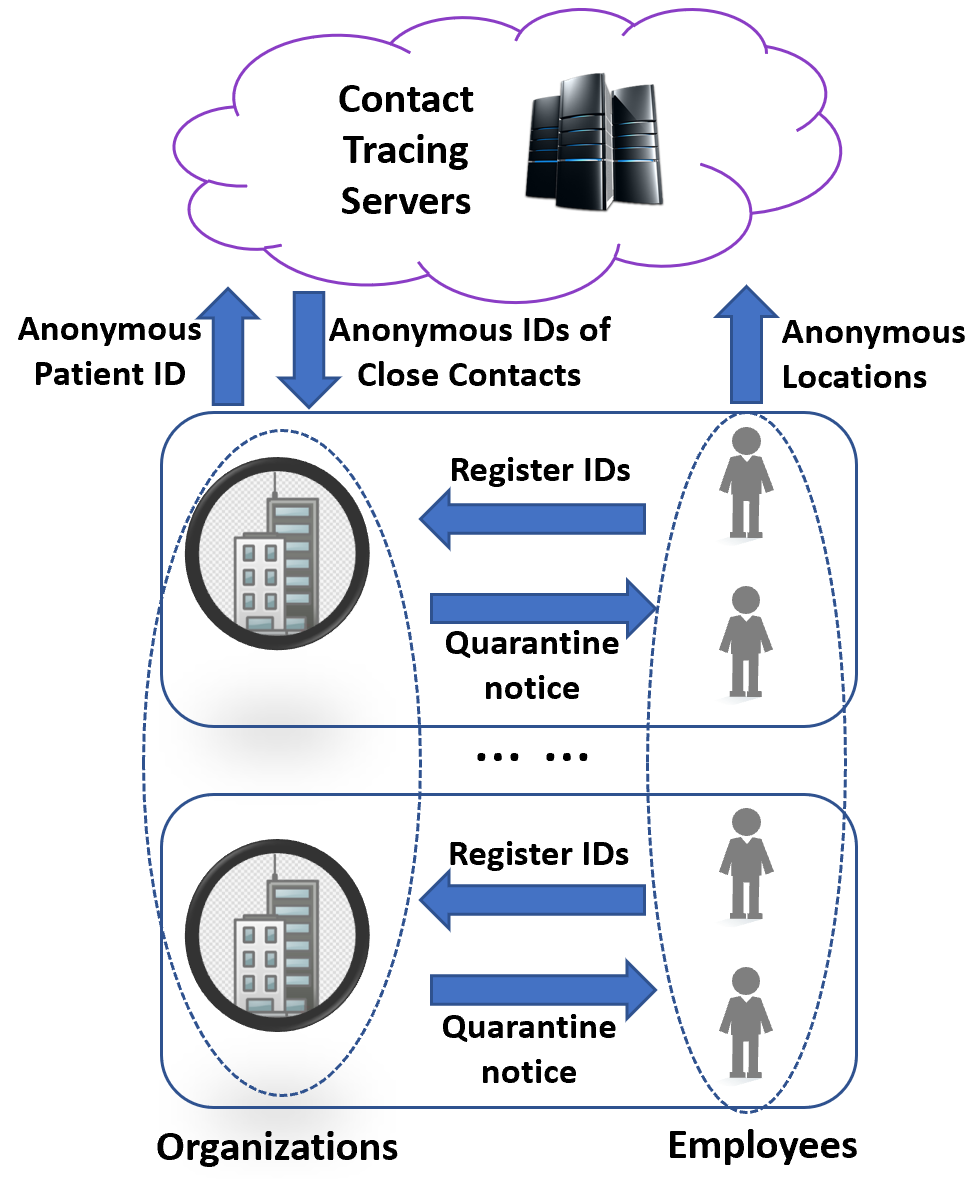, width=0.43\textwidth}
		\caption{An Overview of the PREVENT System}
		\label{fig:system}
	\end{figure}

	\begin{itemize} \itemsep=0pt
		\item We design a novel architecture for privacy-preserving contact tracing and secret sharing based information management protocols. Our system goes above and beyond existing approaches, providing more stringent privacy guarantees for users. No single party in our system will gain access to any information beyond what they already possess, ensuring maximum privacy for all involved.
		
		\item We design highly efficient query algorithms which are able to  identify affected people in large-scale datasets within milliseconds. This is attributed to a unique pyramidal data structure that organizes encrypted user location information at various levels of spatial granularity. With the aid of this data structure, our algorithm can easily  handle the transitive effect, where a person may have been exposed to someone who later developed symptoms after encountering patient zero. This feature allows for more comprehensive and effective contact tracing.
		
		\item Our system can be integrated with a combination of various location gathering methods which include but are not limited to GPS, Bluetooth, QR code, and door swiping systems.
			
		\item  We have designed our system to support multiple organizations, cities, and states, making it scalable and adaptable to various settings. To evaluate our system, we implemented a fully functional prototype and conducted extensive experiments. Our experimental results show that our approach introduces minimal overhead in terms of data management compared to non-privacy-preserving approaches. Furthermore, our query algorithms are as fast as non-privacy-preserving approaches, ensuring that our system can be deployed efficiently in real-world scenarios without sacrificing performance.

	\end{itemize}
	
	The rest of the paper is organized as follows.  Section \ref{sec:related} reviews the related work on privacy-preserving contact tracing. Section \ref{sec:system} presents our proposed PREVENT system. Section \ref{sec:security} analyzes the privacy properties of our system.  Section \ref{sec:exp} reports experimental results. Finally, Section \ref{sec:con} concludes the paper.
	
	\section{Related Works} \label{sec:related}
	
	There have been various works on privacy-preserving contact tracing \cite{tang2020privacypreserving} ever since the pandemic started. However, to the best of our knowledge,  none of them handles the same comprehensive scenarios  or achieves the same strict security goals as presented in our work, and none of them addresses the scalability concerns.
		
	Most of the existing works leverage  short-range wireless technology such as WiFi and Bluetooth to detect human-to-human contact. These approaches have a common limitation. They will miss indirect contacts when another person encounters the lingering virus  after an infected person left the area. Due to the inherent design of storing the direct encountering information locally at users' devices, this line of approaches is hard to be extended  to identify indirect contacts or perform contact tracing queries on a comprehensive trajectory composed of location data collected via various means (e.g., GPS, QR codes). An early example of these works is the EPIC system proposed by Altuwaiyan et al. \cite{EPIC}.  Their server calculates  a weight-based matching score between users based on encrypted connection signals. The encryption scheme adopted in this approach is inherently computationally expensive,  and their experiments only tested 5 pairs of people. More recently, Trieu et al.  \cite{trieu2020epione} propose to use Bluetooth to allow users to exchange and store a randomly generated ``contact token" when they are close to each other. If a user has been diagnosed positively, the user will inform the server to broadcast a set of tokens that are used by him/her. Other users then compare the tokens received from the server to the tokens gathered from contacts to see if they may be exposed. Following a similar idea, Pinkas and Ronen \cite{Hashomer} propose a Hashomer system that also relies on  Bluetooth to detect close contact among users,   record the pseudo IDs of encounters in the application, and let health bureau to broadcast reported IDs of patients to all the users.


	In order to provide users some more control of their privacy, Song et al.\cite{PpCT} propose a notion of self sovereign identity which allows individuals to determine when and whether to share their identities when encountering others. To further improve the privacy protection, some approaches remove the central server and leverage blockchain techniques. For example, Ahmed et al. \cite{ahmed2021dimy} propose that people who have been diagnosed positively can  choose to upload their pseudo IDs to the blockchain. Then, other users can query the blockchain  to check if they have been in contact with the patients. This approach  requires users to actively and constantly check the blockchain which is inevitably time and energy consuming, while our system will automatically notify only the users who are at risk.

	It is important to note that the aforementioned human-to-human contact-based approaches may not be energy-efficient in large-scale settings without sacrificing users' location privacy. In cases where the server is unaware of a patient's region (e.g., city), broadcasting the patient's pseudo IDs to a large number of users throughout the country for self-checking may result in the unnecessary consumption of phone battery for those who are geographically distant from the patient. To narrow down the range of users to be notified, the patient must be willing to surrender some location privacy by providing information such as the cities they have visited. Our proposed approach, in contrast to existing solutions, achieves both scalability and efficiency. Our system can handle users from across the globe without requiring any level of location information from them. The number of messages broadcast is limited to only a few subscribers, and notifications are sent only to users who may have been exposed to the virus.
			
		
	In order to capture indirect contacts between users, some QR code based approaches have been proposed \cite{QRcode}. Such approach requires event owners to set up the QR codes so that users can scan the codes to record places they have visited. Later, they require the patient to report to the event owner about his/her diagnose so that the event owner (or the server) can broadcast those risky places to event participants. That means the server will know all the places that the patient has been to. Moreover, the tracing  is limited to  only places with QR codes.

	There are also some approaches which allow servers to conduct privacy preserving queries on collected  entire trajectories rather than just encounters recorded by short-range communication.   For example, Kim et al. \cite{plosone} propose to use functional encryption to encrypt users' trajectories and then perform queries directly on encrypted data. However, their settings will require all the users to use the same encryption key to generate encrypted trajectories which will be stored by the server. This may not be secure enough since an attacker just needs to compromise a single user to decrypt the whole dataset. Reichert el al. \cite{cryptoeprint:2020:375} propose a theoretical approach that applies secure multi-party computations among all users. As it requires all users to participate in SMC  to calculate if their locations were ever in the infectious area of others, it is extremely computationally expensive  when there are a large number of users like the city and multi-organization setting in our work. Most recently, Zhang et al. \cite{5G} propose a block-chain based scheme to achieve  privacy-preserving contact tracing in 5G-integrated environment. Their system consists of a trusted medical center and fog nodes. Fog nodes are responsible to log the locations of people near them using 5G and store them on a ``permissioned blockchain". Users use their phones to  upload their encrypted identities when passing by checkpoints. The users are also able to check if their routes have included any potentially dangerous locations by checking the blockchain. In their system, users will not know others' exact location information, but the medical center has access to everyone's locations. This is different from our system as we ensure that not any single party, including servers and medical centers, will be able to gain the location information of a user or a patient. 
	
	When it comes to privacy preservation, one may also think about homomorphic encryption and differential privacy. However, these technique may not be suitable here. Consider the number of people and the places they will visit during several weeks. The amount of location information to be analyzed is in astoundingly  large scale. However, homomorphic encryption incurs high computational overhead \cite{Yi2014} and has not been successfully employed for real-time large-scale data set analysis yet. While differential privacy techniques \cite{abdelraheem2018differentially} are effective at identifying general movement patterns in location data while preserving individual privacy, they may not be sufficient to meet the accuracy requirements of contact tracing queries.

	
	In addition, there have been works on secure cloud data storage and retrieval \cite{PrivacyPreservingStorage}. However, those approaches are not  applicable to contact tracing  because they only allow data owners to securely retrieve their own files whereas contact tracing requires  to query on other people's information.

	\section{The Proposed PREVENT System} \label{sec:system}
	
	In this section, we present our proposed privacy-preserving contact tracing system, namely PREVENT (Privacy pREserving Virus ENcountering Tracking).  Our system is designed to take location data recorded via any means by users. It can support large-scale data storage and queries for multiple organizations, multiple cities, states, and even countries. It  consists of three main modules: (i) Privacy-preserving data transmission from users to the servers; (ii) Privacy-preserving data storage at the servers; (iii)   Privacy-preserving multi-generation contact tracing queries (Definition 1). 


\begin{def} \label{def:query}
	{\bf Definition 1: Privacy-preserving Multi-Generation Contact Tracing Query}: {\em Let $D$ be the infectious distance, $\tau$ be the infectious time window, and $T$ be the incubation period. Given a patient's pseudo ID $u_0$, the privacy-preserving contact tracing query $Q$ returns a list of pseudo IDs $U=\{u_1, ..., u_k\}$, whereby  $u_i$  satisfies the following condition: at least one location of $u_i$ appears within $D$ distance to at least one location of the previous identified close contact $u_j$ during $T$ and $t_{loc_i} - t_{loc_j}\leq\tau$.}
\end{def}

In what follows, we first present the threat model and then elaborate the detailed algorithms for each module.

\subsection{Threat Model}
	
There are three parties in the PREVENT system: contact tracing service provider (servers), subscribers, and users. The security assumptions regarding each party is summarized as follows. 
	
Users are people who are willing to provide their location information anonymously to contribute to the contact tracing to help the society. They  will  submit their location information in an encrypted form at the end of each day to the servers. The location information can be collected in a variety of means including but not limited to  GPS, Bluetooth, door swiping systems, etc., to record different types of locations including static locations like supermarket and dynamic locations like buses. The detailed location reporting process will be elaborated later. 
		
Subscribers could be  organizations, companies, medical providers, government agencies,  etc. They are assumed to  faithfully  carry out assigned identity verification tasks for users.  Only subscribers but not users are allowed to launch contract tracing queries. Users can choose to register their real identities with any subscriber, and users decide whether  to inform the subscribers about their diagnosis or not.
		
The contact tracing service provider is assumed to employ a distributed server architecture, where each server operates independently of the others and does not collude, unless simultaneously compromised by attackers. The real-world deployment of multiple servers can be accomplished by leveraging third-party login mechanisms that are prevalent today. For instance, various web service providers offer their users the ability to authenticate using Google or LinkedIn accounts. In this context, a PREVENT server can be located at the main contact tracing service provider, while others can be deployed at Google or LinkedIn. Rather than providing authentication services for the primary service provider, servers at Google and LinkedIn can execute security protocols for the PREVENT system.
			
In our system,  we guarantee that no single party except the user him/herself knows the exact locations. The privacy goals with respect to each participating party are summarized as follows.
\begin{itemize} \itemsep=0pt
	\item \textbf{Users are fully anonymous to service providers.} Servers in the PREVENT system will not know the plain texts of users' real identities and locations.
	
	\item \textbf{Users' trajectories are fully anonymous to subscribers.} The system subscribers, such as health care providers, will not have access to any individual's location data, which includes their whereabouts and timestamps. If an individual needs to be quarantined, only their identity is communicated to the relevant subscribers, who are responsible for disseminating notifications to these users. The notification message does not contain any specific information about the time and place where a user may have been exposed to the virus. Rather, it simply notifies the users that they may have been exposed to the virus and are potentially at risk.

	\item \textbf{Peer users do not know each other's anonymous ID or reported locations}: Users even from the same organization will not know each other's anonymous ID and what location information that others have reported.
	
\end{itemize}

To deploy our system in the real world, it should be integrated with existing security measures that encompass authentication, anti-malware, and network communication security. To illustrate, user accounts should undergo validation and be authenticated to their respective mobile applications by the subscribers. Additionally, mobile devices used by the users must be safeguarded by anti-malware software. As there are numerous established techniques that address the common security concerns, and given our primary focus is on privacy preservation, we refrain from delving into the specifics of such techniques in this context.

It is worth noting that our system has been designed to withstand potential misuse by users. Firstly, users are prohibited from initiating contact tracing queries, which prevents malicious actors from flooding the servers with fraudulent requests. In the event of a malicious user submitting a false diagnosis to the subscriber, it would only result in a false alarm being triggered for other users. It would be infeasible for the malicious user to submit multiple false diagnoses within a short period of time, as this would be noticed by the subscriber. Secondly, in the event of a user deliberately providing incorrect location information, it would only affect the user in question, as they would not receive appropriate notifications about potential virus exposure. Moreover, our system design also prevents a user from submitting a large volume of fake location data since each location data needs to be associated with a pseudo ID issued by the subscriber. The number of valid pseudo IDs a user has will not exceed  a reasonable threshold of the number of locations a person could feasibly visit in a day. If the malicious participant reuses the same pseudo ID for multiple locations, that can be easily detected by the server. 

Finally,  all users who participate in our system do so voluntarily. It is ultimately their own decision whether to disclose their diagnosis to others. Our system has not been devised with the intention of compelling users to report their locations or diagnosis. As a result, inaccurate contact tracing resulting from users who are unwilling to report their locations is not considered as a metric for evaluating the efficacy of our system.

			\begin{figure*}[!t]
		\centering
		\epsfig{file=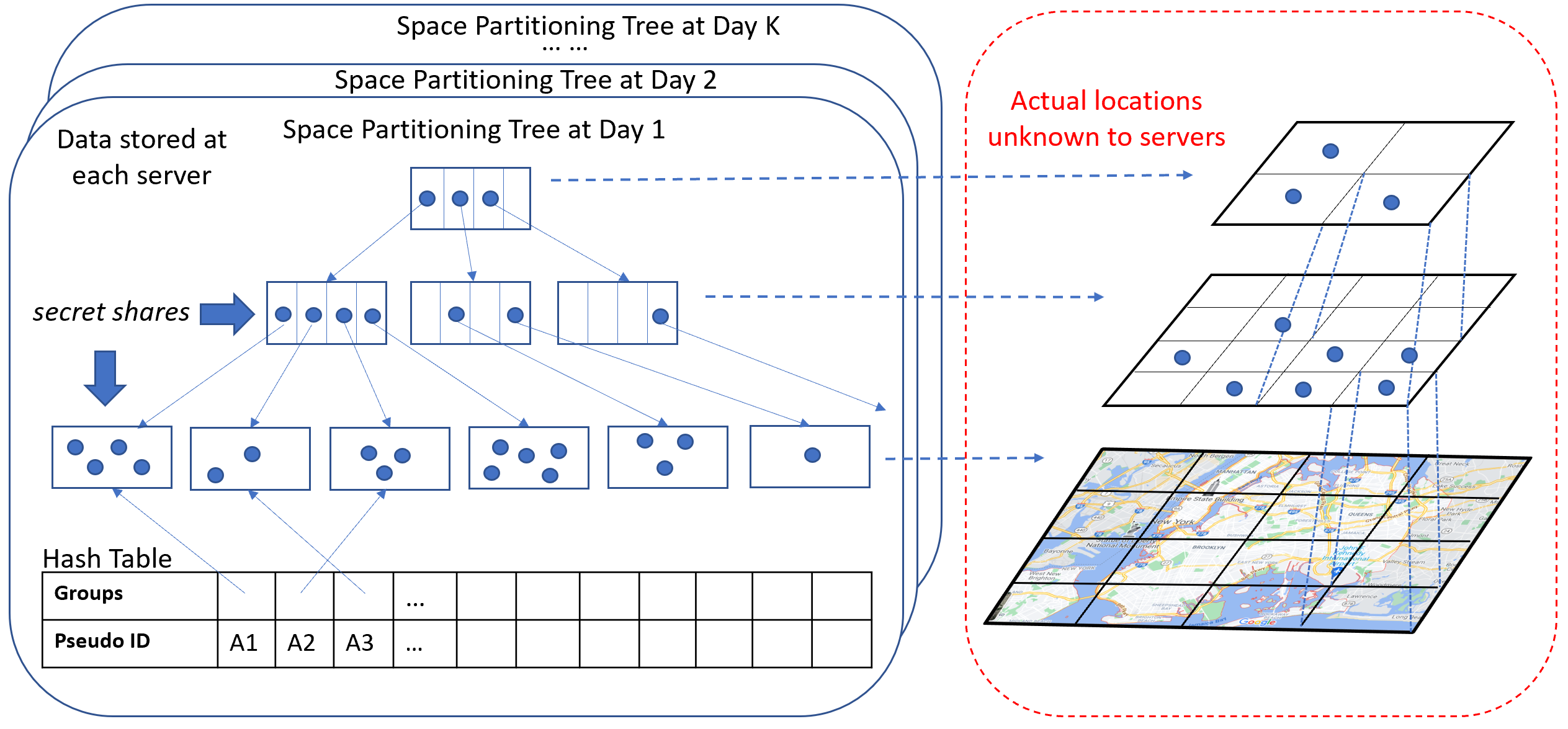,width=5.5in}
		\caption{An Overview of Data Structure}
		\label{fig:datastructure}
	\end{figure*}
	
	
\subsection{Privacy-preserving Data Transmission}
	
Users who intend to partake in the contact tracing program need to complete the registration process through designated subscribers.  Note that each user only needs to register with one subscriber and the subscriber is not necessary to be located in the same city as the user. A further noteworthy feature of our system is that even if a traveling user comes into contact with a patient in a location outside the traveling user's own city or the subscriber's city, the traveling user will still receive notifications.

The subscribers will provide a set of pseudo IDs for the users to interact with the contact tracing server later on through their mobile apps. The PREVENT mobile app at the user side will perform data anonymization that not only hides user's location information but also enables contact tracing queries to be computed directly on such anonymized data. Specifically, the mobile app performs two major functions: (i) stay point recording; and (ii) location reporting.  

Stay points, i.e., the location that the user stays longer than the infectious time window $\tau$, can be detected by many  existing means of location detection. For example, GPS can be used to record the locations where a user lingers longer than $\tau$, such as a supermarket, a restaurant and a shopping mall. Bluetooth technology similar to some existing apps \cite{Hashomer} can be used to activate the location recording when the user encounter other people who rode the same car, bus, subway, or airplane. Specifically, once a human subject is detected by the Bluetooth in the user's smart phone, our app will start recording the duration of the encounter. Once the encounter lasts more than the infectious time, the user's current GPS location will be recorded to be submitted to the server at the end of the day. The encounter with the same person during the same time period will only trigger one-time location recording. Similarly, locations collected from the door swiping systems will be recorded by our app for reporting only if the user stays in the room long enough. To unify the representation of locations, any non-GPS location such as those obtained from indoor positioning systems or QR codes will be converted into coordinates based on where the location detection systems are located. 

With regard to location reporting, our mobile app employs the additive secret share scheme to partition each recorded stay point and corresponding timestamp into secret shares. Different pseudo IDs received from the user's subscriber will be associated with secret shares of different stay points. This approach ensures that neither the servers nor any third parties have knowledge of the user's real ID or precise location. Moreover, the additive secret share scheme offers the additional benefit of ensuring that the user's location remains confidential unless an attacker controls more than $k$ servers, which comes at a high cost to the attacker. To further prevent servers from learning the sequence in which locations were visited, the user's mobile app will not immediately report the user's current location to the servers. Instead, users will only need to submit the set of location updates once at the end of each day. Since the timestamps of each location also constitute a piece of secret share, the servers will be unable to discern the order in which locations were visited. Specifically, at the end of each day, the mobile app will encrypt the secret shares of all stay points along with their associated pseudo IDs using the servers' public keys and send them to the servers. The secret shares of the same stay point will be sent to different servers, and the secret shares of different stay points will use different IP addresses, which can be achieved using VPN apps. This method will help to prevent servers from linking multiple location reports to the same user.
	
Storing the secret shares of users' location information fulfills the first design goal -- the privacy protection. It is still not efficient for the subsequent large-scale contact tracing queries. This is because a brute force approach to finding people at risk would be to compare all of the patient's locations with those of all other users', which is obviously time consuming  especially when these comparisons need to be performed via secure computation protocols on secret shares. Therefore, we further enhance the user data organization and develop a data filtering stage to significantly narrow down the search space and obtain a much smaller set of candidate location sets for fine-grained analysis.

Our idea is to hierarchically partition the overall space under consideration into grids with equally-sized cells as shown in Figure \ref{fig:datastructure}. For ease of the calculation, we use the minimum bounding square for the overall space. The minimum latitude and longitude are denoted as  $x_0$ and  $y_0$, respectively. The number of levels in the space partitioning  tree is denoted as $H$,  the width of the grid cell at the i$^{th}$ level is denoted as $w_i$ where the first level is the lowest. The height and widths of the grid cells at each level are known to all parties, thus the mobile app at each user side can automatically calculate which grid cells the user is currently located using Equation \ref{eq:gid}, where $x$ and $y$ are the latitude and longitude of the user's location. The grid IDs (denoted as GIDs) will be  split into secret shares and appended to the previously generated secret shares of the exact location and timestamp.

\begin{equation}\label{eq:gid}
	GID_i(x,y) = \lfloor\frac{x-x_0}{w_i}\rfloor + \lfloor(\frac{y-y_0}{w_i}-1)\cdot\frac{W}{w_i}\rfloor
\end{equation}
	
\begin{figure}[!t]
		\centering
		\epsfig{file=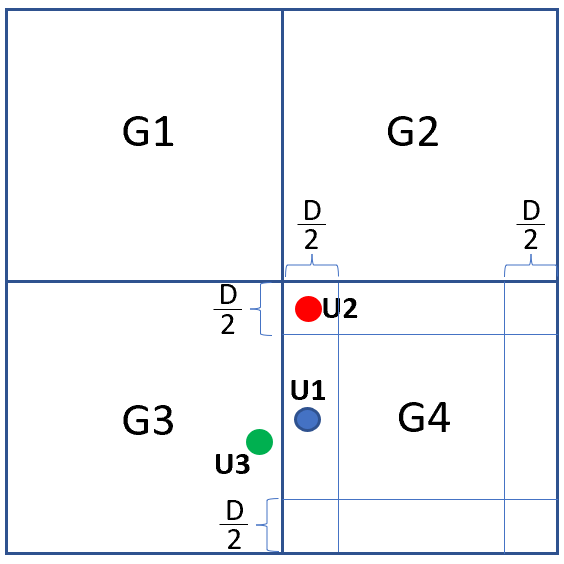,width=0.21\textwidth}
		\caption{Special Handling of Locations near Borders}
		\label{fig:border}
	\end{figure}

There is a special handling of user's locations within $\frac{D}{2}$ distance to the border of a grid cell as illustrated in Figure \ref{fig:border}. In addition to the previous message, we will create one to three additional messages that contain the new secret shares of the location and the neighboring grid cell IDs. Figure \ref{fig:border} shows an example. The colored circles represent locations of users $u_1$, $u_2$ and $u_3$, respectively. We can see that $u_1$ is located less than $D$ (infectious distance) to the border of grid cell $G_3$, which means some users such as $u_3$ in grid cell $G_3$ may be within the infectious range of $u_1$. In order to allow  the subsequent contact tracing query to be efficiently conducted within a single grid cell, we will let the servers store  $u_1$'s pseudo ID in the grid cell $G_3$ as well. Specifically, a message that contains newly generated secret shares of $u_1$'s location, grid cell $G_3$ and its parent grid cell IDs will be sent to the servers in addition to the  message for $u_1$'s original grid cell $G_4$. Similarly, $u_3$ will also be stored in grid $G_4$'s group at the server side. As for the corner case like $u_2$, three additional messages will be created to include  $u_2$ in grid cells $G_1$, $G_2$ and $G_3$.   

 \begin{algorithm} [!t]
	\caption{Location Data Transmission from a  User}\label{alg:shares}
	\begin{algorithmic}[1]
		\STATE Collect a set of pseudo IDs from the server $UID^{u}$=$\{UID^{u}_1, UID^{u}_2, ..., UID^u_m\}$ (once at the first usage)
		\STATE Establish a sufficiently large prime number $Q$
		\FOR{each stay point $loc_k$ at time $t$ of the day}
		
		\STATE \textbf{Offset Coordinates To Be In Range [0,360]}
		\STATE $long \gets OffsetCords(loc_k.longitude)$
		\STATE $lat \gets OffsetCords(loc_k.latitude)$
		\STATE
		\STATE \textbf{Calculate $N$ Secret Shares Of $loc_k$}
		\FOR{$i = 1; i < N; i++$}
		\STATE $X_i \gets Random Integer(0,Q)$
		\STATE $Y_i \gets Random Integer(0,Q)$
		\ENDFOR
		\STATE $X_N \gets (long - \sum_{i=1}^{N-1} X_i) \mod Q$
		\STATE $Y_N \gets (lat - \sum_{i=1}^{N-1} Y_i) \mod Q$
		\FOR{$i = 1; i \leq N; i++$}
		\STATE $l_i \gets (X_i, Y_i)$
		\ENDFOR
		\STATE
		\STATE \textbf{Calculate The Ids For The Insertion Tree}
		\FOR{lev = 2; lev $\leq$ H; lev ++}
		\STATE Calculate $GID_{lev}$ of $loc_k$
		\STATE Calculate $N$ secret shares of $GID_{lev}$
		\FOR{$i = 1; i < N; i++$}
		\STATE $G_{lev}^i \gets Random Integer(0,Q)$
		\ENDFOR
		\STATE $G_{lev}^N \gets (GID_{lev} - \sum_{i=1}^{N-1} G_{lev}^i) \mod Q$
		\ENDFOR
		\STATE
		\STATE \textbf{Calculate $N$ secret shares of timestamp $t$}
		\FOR{$i = 1; i \leq N; i++$}
		\STATE Send $\langle UID^u_k, t_i, l_i, G_1^i, G_2^i, ..., G_{H}^i\rangle$ to Server $i$
		\ENDFOR
		\ENDFOR
	\end{algorithmic}
\end{algorithm}
	
This example also hints that the size of the grid cell at the lowest level should not be too small. In the extreme case when it is smaller than the infectious distance as the overlap from each cell is half the infection distance, all the locations in the cell will need to be included in some neighboring cells. Therefore, we set the cell size at the lowest level to at least 2D (i.e., two times of the infectious distance).  It is worth noting that even though the location at the border of edge cell produces a couple of more GIDs, the servers will not be able to know whether the user is located at the edge of a cell based on the total number of GIDs the user sent. This is because the GIDs of all the locations are sent together at the end of day, and different users may visit different numbers of locations. It is indistinguishable to the servers whether a longer list of GIDs is caused simply by users visiting more places or users located at the border of cells. Algorithm \ref{alg:shares} summarizes operations conducted by  the user's mobile app.
	

\subsection{Privacy-preserving Data Storage}
At the server side, each server maintains a hash table and a space partitioning tree everyday for latest $T$ days, where $T$ is the incubation period. Such storage can significantly improve the efficiency of the subsequent contact tracing queries as discussed in the next subsections. The hash table stores pseudo IDs, the secret shares of locations, timestamps, and grid IDs along with a pointer to the leaf node of the space hierarchy.

In the space partitioning tree (left side of Figure \ref{fig:datastructure}), each leaf node stores secret shares belonging to users located within the same grid cell at the lowest level of space partitioning. However, the server is unaware of which leaf node corresponds to a particular physical grid cell. From the server's perspective, it only sees multiple hierarchically organized groups of secret shares.  An internal node is composed of multiple entries, with each entry storing a single secret share of the higher-level grid cell ID of the first user inserted into that cell. 
The server still does not possess knowledge of the exact grid cell that the internal node represents. The number of entries in each node may be fewer than the total number of grid cells based on the space partitioning since grid cells that have not been visited by users will not have any representation in the tree.  Also, attributed to the property of the secret sharing scheme, each user's secret share is different from one another even if they are in the same grid cell, making it impossible for a single server to infer the exact grid cell location that  a group of secret shares belong to. 

Given a user's location update which consists of  secret shares of  a set of locations (or stay points) visited by the user in a day, the server will insert each piece of location information as follows.  Starting from the root level of the space partitioning tree,  the servers will conduct a collaborative  protocol to compare the GID secret shares of the newly reported location with  those of the representative user at the root level in the current space partitioning tree. If a match is found, the secure comparison continues to the children nodes of the matching GID. This comparison process continues until the leaf node of the space hierarchy is reached. If there is no matching GID at any level, a new entry will be created to store the user's pseudo ID, and the user becomes the representative user of this entry. The exact GID is never revealed to any server at any level of the process.

There are four main steps to securely compare the secret share of the grid ID in the new message (denoted as $G_{ui}$) and the grid ID of the representative user (denoted as $G_i$) in the space partitioning tree at Server $S_i$.  First, server $S_i$ calculates $G_{ui} - G_i$, and stores this difference in $d_i$. Second, all the servers execute the secure random number generation protocol \cite{Keller20} to  generate a random number $r$. At the end of this protocol, each server only has a share $r_i$ of this random number $r$ but does not know the value of $r$. Then, each server applies the secure multiplication protocol \cite{Keller20} to derive $v_i=d_i \cdot r_i$, and shares $v_i$ with all the other servers. Finally, every server calculates the sum of $v_i$s, i.e., $\sum_{i=1}^{N}v_i$. If this sum is zero, that means the two grid cell IDs match, and each server will insert the secret share of the new position into the corresponding entry in its own space partitioning tree. During the whole process, each server never sees the exact value of the grid cell IDs being compared. 

%
%

\begin{figure}[!t]
	\centering
	\epsfig{file=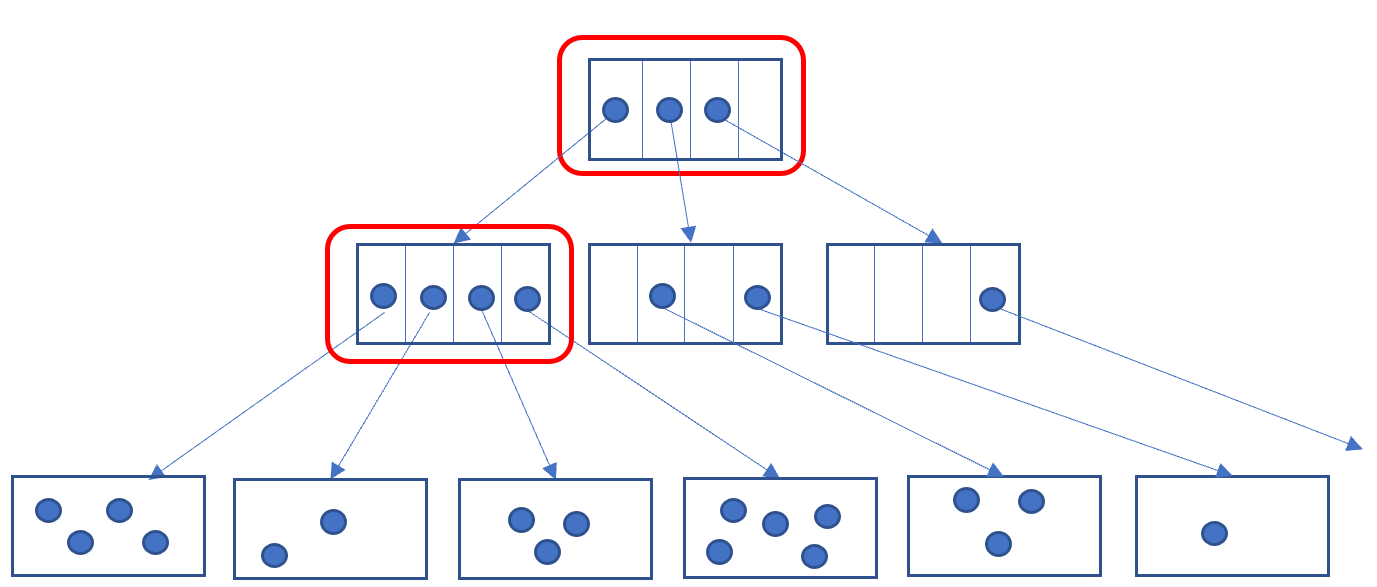,width=0.4\textwidth}
	\caption{Comparisons Needed for Inserting a Location}
	\label{fig:insertcost}
\end{figure}

\subsection{Privacy-preserving  Multi-generation Contact Tracing Queries}

As shown in Definition 1, the goal of a contact tracing query is to identify users who had visited the same places as the patient during the infectious time window. Depending on the quality of the collected location data, the query results may contain false positives. For instance,  users with nearby coordinates provided by indoor positioning systems may be separated by a physical wall, and thus they may not infect each other. Such false alarms will not harm public health. It is more crucial not to miss individuals at risk, and our query algorithm will guarantee that there is no false negatives  based on the collected location data.

Since each server possesses only a piece of secret share of the user's location data, the contact tracing queries will be conducted via a collaborative protocols among multiple servers without leaking users' location information to any individual  server.

\begin{figure}[!t]
	\centering
	\epsfig{file=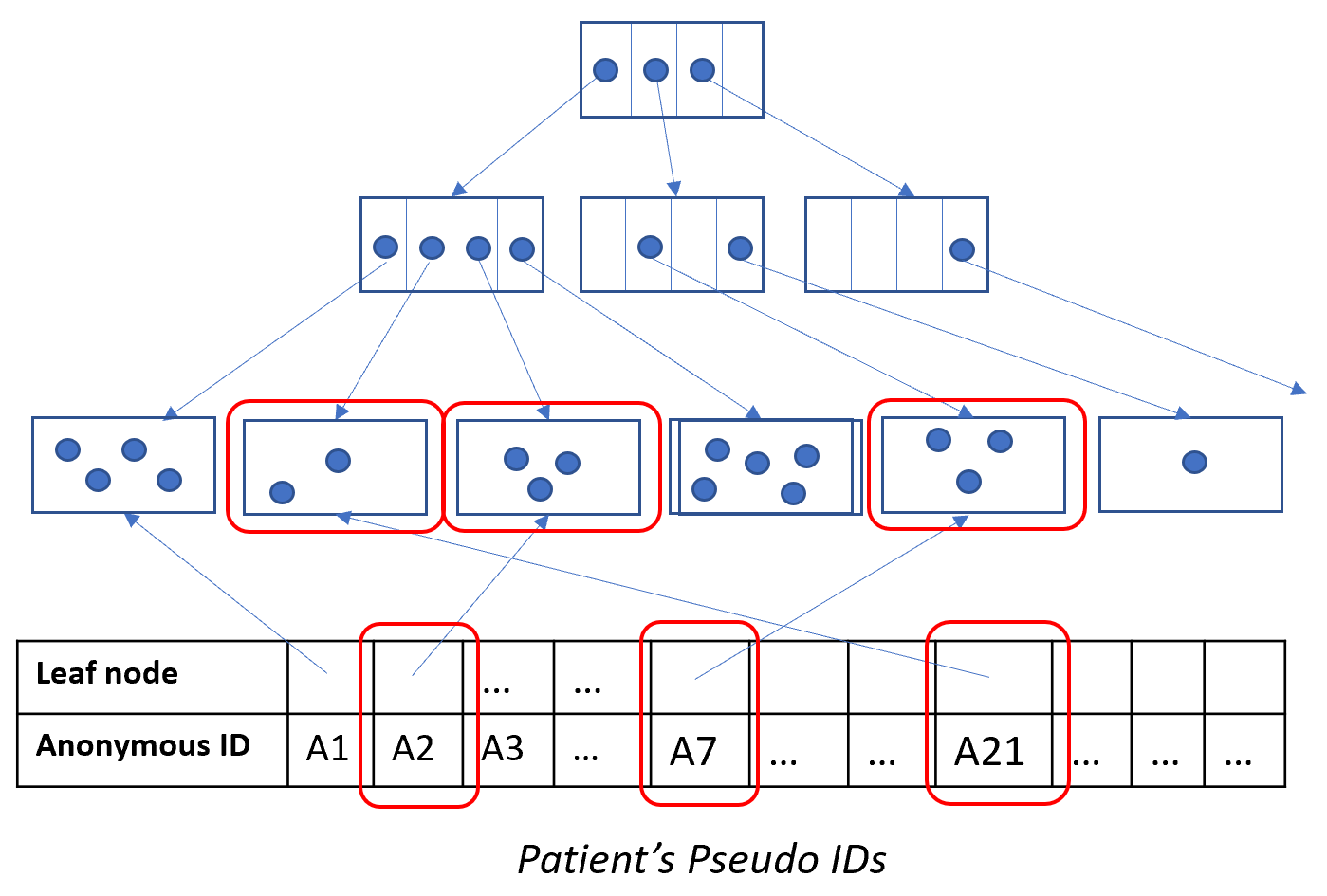,width=0.5\textwidth}
	\caption{Comparisons Needed for a Query }
	\label{fig:querycost}
\end{figure}


Once a user has been diagnosed, he/she may inform the subscriber (e.g., medical provider) where  he/she registered. The subscriber will then send a contact tracing query to the server. The query  will consist of the set of pseudo IDs of the user. From the server's point of view, the server does not know whether the received pseudo IDs belonging to the same or different users. The server runs the same query protocol for each received pseudo ID as follows.  First, each server retrieves the stored  location share  corresponding to the pseudo ID.   Then, all the servers collaboratively and securely compare the location visited by the patient with those of  other users  to see if they are within the infectious region and time window. The pseudo IDs of the identified users will be treated as new query inputs, and the query process will be repeated until all possible contacts are identified within the incubation period of the first patient. Finally, the main contact tracing server broadcasts the retrieved pseudo IDs to all the subscribers which will inform the corresponding users with a simple message that they may have been exposed to virus.  

We now proceed to elaborate the secure comparison of location shares among servers. Assume there are $n$ servers $S_1$, ..., $S_n$, each of which contains secret shares of the users' location information during the length of the incubation period (e.g., last 14 days). Given a patient's pseudo ID $UID_p$, each server will use the hash table to retrieve the leaf node in its space partitioning tree that the patient's location belongs to as shown in Figure \ref{fig:querycost}. As our data transmission algorithm (in Section 3.2) ensures that query results regarding a particular location will be inside the same gride cell (i.e., leaf node) that the patient's location resides, the servers just need to compare locations within each retrieved leaf node. The structure of the space partitioning tree is the same across all the servers, i.e., the users are grouped in the same way. The only difference among the space partitioning trees is that different servers store different parts of the secret shares of the same location. 

Specifically, let $l_i(x_i, y_i)$ denote the secret share of the location of the user who is in the same group as the patient at server $S_i$, and $l_{iq}(x_{iq}, y_{iq})$ denote the querying location (i.e., the patient's location). The goal of this protocol is to check if the user is within the distance of $D$ of the patient. First, each server computes the differences between the secret shares of the user's and the patient's x and y coordinates:  $\Delta x_i=x_i-x_{iq}$, $\Delta y_i=y_i-y_{iq}$. Then, all the servers together execute the secure multiplication protocol \cite{Keller20} to compute the square of the differences, i.e., $d_{xi} = (\Delta x_i)^2$, $d_{yi} = (\Delta y_i)^2$. Next, each server sum up $d_{xi}$ and $d_{yi}$ to obtain the share of the square of the  Euclidean distance between the user and the patient, denoted as $d_i$. Finally,  all the servers perform the secure comparison protocol to check if the square of the Euclidean distance is smaller than $D^2$ (the square of the infectious distance). The protocol is outlined in Algorithm \ref{alg:query}. 

\begin{algorithm} [!t]
	\caption{Securely Compare User's and Patient's Locations}\label{alg:query}
\textbf{Input:} ($S_i$, $l_i$, $l_{iq}$, $D$)\\
\textbf{Output:} $\Delta l \leq D$
	\begin{algorithmic}[1]
		\REQUIRE $D$ is the infection distance squared
		\REQUIRE $l_i \rightarrow (X_i,Y_i)$
		\REQUIRE $l_{iq} \rightarrow (X_{iq}, Y_{iq})$
		\FORALL{$S_i$}
		\STATE $\Delta X_i \gets X_i - X_{iq}$
		\STATE $\Delta Y_i \gets Y_i - Y_{iq}$
		\STATE $d_{xi} \gets SecureMultiplication(\Delta X_i, \Delta X_i)$
		\STATE $d_{yi} \gets SecureMultiplication(\Delta Y_i, \Delta Y_i)$
		\STATE $d_i \gets d_{xi} + d_{yi}$
		\STATE $SecureComparison(d_i^2, D^2)$
		\ENDFOR
		
	\end{algorithmic}
\end{algorithm}

The cost of finding close contacts of the patient can be estimated as follows. Suppose that there are currently  $N_u$ users' information in the system, and the average number of locations recorded for each user is $\kappa$. Assuming all the users' locations are uniformly distributed in the space, each grid cell at the lowest level will contain approximately $\frac{N_u \cdot \kappa}{(\frac{W}{w_1})^2}$ locations. If the average length of users' trajectories is $L_u$, the average number of grid cells that a user falls in can be estimated as $\frac{l_u}{w_1}$. Since the query will compare the patient's location with other users in the same grid cell, the query cost can be estimated as the product of the number of grid cells visited by the patient and the number of locations in each grid cell as shown in  Equation \ref{eq:querycost}. The cost of the multi-generation query is simply the sum of these individual query cost. 

\begin{eqnarray}\label{eq:querycost}
	C_{query} = \frac{N_u \cdot \kappa}{(\frac{W}{w_1})^2} \cdot \frac{l_u}{w_1}
	=  \frac{N_u\cdot \kappa \cdot l_u \cdot w_1}{W^2}
\end{eqnarray}

\subsection{Choice of Space Partitioning}

From Equation \ref{eq:querycost}, we can observe that the query cost increases with the grid cell size $w_1$. This is because larger grid cells contain more locations to be compared. We can also see that the cost reaches highest when $w_1$ equals the space width $W$ (i.e., no space partitioning). In general, the smaller the size of the cells at each level, the fewer number of comparisons for a query but more comparisons during the insertion. In order to find the trade-off point, we analyze the insertion and query cost as follows.

For ease of discussion, we consider only two levels of space partitioning and call the upper level partitioning `regions' and lower level partitioning `grid cells'. The following reasoning can be easily extended to any desired number of levels.  The insertion of a piece of user information involves $N_r$ rounds of comparisons with the region IDs and another $N_g$ rounds of comparisons with the grid IDs in the retrieved region. The total number of comparisons introduced by space partitioning is thus $N_r+N_g$. Therefore, the smaller the value of $N_r$ and $N_g$,  the fewer comparisons would be needed during the insertion.

Next, we look at the impact of space partitioning at the query side. Suppose that there are currently  $N_u$ users' information in the system, and the average number of locations recorded for each user is $x$. Without any space partitioning, a querying trajectory that consists of  $y$ locations will need to be securely compared with each location in the $N_u$ users' trajectories, which will results in $N_u\cdot x\cdot y$ comparisons. The number of users is typically very large, i.e., $N_u$ is very large. Such brute force comparisons would be inevitably computationally expensive.

With the space partitioning tree in place, the number of needed comparisons for a query can be drastically reduced. Specifically, we can first locate the regions and grid cells where the querying trajectory resides. Assume a trajectory contains $\lambda$ locations, for each location point, we need to locate its corresponding region and grid cell. The number of secure comparisons for this would be $\lambda\cdot(N_r+N_g)$.  After that, we only need to conduct secure comparisons between the query location and the locations in the retrieved grid cell  instead of the entire dataset of locations.  The number of the comparisons of the exact locations is determined by the total number of users ($N_u$) and the  total number of grid cells ($N_r\cdot N_g$). The cost can be estimated as $\frac{N_u}{N_r\cdot N_g}$ assuming locations are uniformly distributed among grid cells. To sum up, the total number of comparisons for a contact tracing query is shown in Equation \ref{eq:cost}.
\begin{equation} \label{eq:cost}
C_q = \lambda\cdot(N_r + N_g + \frac{N_u}{N_r\cdot N_g})
\end{equation}
For $C_q$ to reach minimal,  $N_r$ and $N_g$ and $\frac{N_u}{N_r\cdot N_g}$ should be equivalent to one another, which leads to the following equation.
\begin{equation} \label{eq:ng}
N_r = N_g = \sqrt[3]{N_u}
\end{equation}
The following simple example illustrates the orders of magnitude computational savings with the aid of the space partitioning.  Assume that there are total 100M uniformly distributed location points. Without space partitioning, a query trajectory with 10 location points will need 10$\times$100M=10$^9$ secure comparisons. If the overall space under consideration is divided into 464 regions and each region is divided into 464 grid cells according to Equation \ref{eq:ng}, each cell will contain about $\frac{10^8}{464\cdot 464}$ = 464 location points.  The total number of secure comparisons would be 10$\cdot$(464 + 464 + 464) = 13920, which is $\frac{10^9}{13920} \approx$  71,839 times less than the case without space partitioning. When there are total 1B location points, the gain from space partitioning would be even more significant, i.e., $\frac{10^{10}}{10^4}=10^6$.  Note that the determination of the size of the grid cell should also consider the infectious distance in that the size of a grid cell should be large enough to contain users who fall in the infectious region with respect to the querying trajectory. The minimum size of a cell should be larger than the infection distance $D$.

In  the real world scenario,  retrieving the group of people who have been in close proximity of the patient would be more urgent than the insertion of a user's location data. Thus, we propose to give the query optimization higher priority when determining the granularity of  space partitioning.

\section{Security and Privacy Analysis} \label{sec:security}

There are three types of parties in our system: servers, subscribers, and the people being tracked. Our system ensures that none of these parties gains more information than necessary. This also means that an attacker cannot gain location information by attacking any single party. First, when the server processes a contact tracing query, it does not know the plaintext locations that the patients have been to. After the server finds users who may be close contacts of the patients, it does not have access to the real identities of these contacts or the plaintext locations of their whereabouts. Therefore, our system does not reveal patient or close contact locations to the server. Second, the server only sends the pseudo IDs of close contacts to subscribers, so subscribers do not have access to the exact locations of any patient or the close contacts of the patient. Thus, our system does not reveal locations to subscribers. Third, users who may be at risk receive a simple message stating "you may have been in contact with the virus". From the message, the users cannot determine which patient they were in contact with, or when and where they encountered the patients. As a result, our system does not reveal any patient locations to close contacts. The following are formal definitions of privacy guarantees achieved by our system and their proofs.

\begin{theo} Without any background knowledge, the probability that a server reveals the real identity of a user is no more than $\frac{1}{||D_{pi}||\cdot||D_{ri}||}$, where $D_{pi}$ is the domain of all possible pseudo identities,  $D_{ri}$ is the domain of all possible real identities, and symbol `||D||' denote the number of elements in set $D$.
\end{theo} 

{\em Proof}: Each server has only secret shares of users' pseudo IDs. Since each location is associated with a different pseudo ID and IP address, the server will not know which set of location information belongs to the same user from the location    reporting process. Thus, given an individual secret share, the server may guess the secret share is corresponding to one of all possible pseudo IDs, i.e., $\frac{1}{||D_{pi}||}$. Given a pseudo ID, the chance for the server to correlate it with the real identity without any background knowledge is $\frac{1}{||D_{ri}||}$. By multiplying these two probabilities, we obtain the probability the server may infer the real identity of a user from the received secret share. 

Due to the security of the underlying threshold Shamir secret sharing scheme \cite{shamir1979how},  as long as the number of colluding servers is less than $k$, these servers cannot derive the original data with the probability higher than that stated in Theorem 1. In addition, the equality and secure comparison protocols have proven to be secure and provided by the well-known MP-SPDZ library \cite{Keller20}. As a result, the servers will not learn anything about the underlying values while executing these protocols.   $\blacksquare$

\begin{theo} Without any background knowledge, the probability that a server knows the smallest grid cell (at the leaf level of the partitioning tree) that a user's location belongs to is no more than $\frac{1}{N_g}$, where $N_g$ is the total number of grids at the lowest level of the space partitioning tree. 
\end{theo}

{\em Proof}:  Each server has secret shares of users' location information. Each location secret share is associated with a different pseudo ID and IP address, which prevents the server from correlating multiple locations to the same user.  Given an individual location secret share, the server may at most guess this location is in one of $N_g$ possible grid cells, and thus the location disclosure probability is $\frac{1}{N_g}$.  $\blacksquare$

Note that this probability is very low as $N_g$ is typically very large. For example, a 3-level partitioning tree with 100 sub-partitions in each partition at each level yields total $100\cdot100\cdot100$= 1M cells at the final level, i.e., $N_g$ = 1M. Moreover, the chance the server knows the exact location of a user from the secret share is even lower which will be $\frac{1}{D_l}$, whereby $D_l$ is the domain of all possible locations in the service area. 


\begin{theo} With background knowledge of an outbreak location, the probability that a server correlates the patient's location secret shares with the actual grid cells is no more than $\frac{q!(4N_v^2-q)!}{(4N_v^2)!}$, where $N_v$ is the number of grid cells the patient's trajectory intercepted, $q$ is the total number of patient's reported locations. 
\end{theo}

{\em Proof:} If the server has background knowledge that an outbreak of infections occurred at a subscriber's location and the patient is from this subscriber, the server may assume that this patient was infected at this subscriber's location, and attempt to infer other locations visited by this patient. Note that the server  assumption itself may not hold since our system allows users to register with any subscriber,  even those located in cities different from the users' location. In the following, we proceed with the analysis by following the server's assumption. Let $L_u$ denote the patient's trajectory length and let $w_1$ be the length of the smallest grid cell. The number of grid cells intercepted with the patient's trajectory  can be estimated as $N_v$=$\frac{L_u}{w_1}$. Using the grid cell that the subscriber is located as the center, the patient's trajectory may reach grid cells within the radius of $N_v$ cells. For simplicity, we approximate this total area as a square shape instead of a circle. The number of grid cells in this area will then be $(2N_v)^2$ = $4N_v^2$. Since the server does not know the visiting order of the locations, the first randomly picked location secret share could be mapped to one of $4N_v^2$ grid cells; the second location secret share could be one out of $4N_v^2-1$ remaining grid cells; and so on. There are total $\frac{(4N_v^2)!}{q!(4N_v^2-q)!}$ choices of mapping from the  secret shares of the patient locations to grid cells.  Thus, the probability of inferring the actual grid cells that the patient had passed by is the inverse of the above total choices.  $\blacksquare$

We show that the above probability is very small in the real world. Given a grid cell that is 12m wide, a 5-mile trajectory will intercept about 880 cells, i.e., $N_v=880$. The possible area the patient may visit could contain  $4N_v^2  \approx$3$\times 10^6$ cells. Even if the patient reported only 3 locations (i.e., $q=3$), the location inference probability is still as low as $\frac{1}{(3\times 10^6)\times (3\times 10^6 -1) \times (3\times 10^6-2)}$$\approx\frac{1}{27\times 10^{18}}$.  

Theorem 3 also applies when the server intends to correlate some users' location  secret shares to grid cells based on the knowledge of some hot spots such as a place which just held a fair with a large number of attendants. In fact, inferring grid cells using population density of  grid cells would be even harder than the previous case when the server knows a patient is from a specific subscriber. This is because density of grid cells vary throughout a day. The server only receives aggregated density information (i.e., all the users who visited the place throughout the day)  since the timestamps are hidden. Also, due to the generation of multiple grid IDs for locations at the border of cells, the density of each cell observed by the server already deviates from the real world density. Further, groups of similar densities are indistinguishable from one another. 

\begin{theo} With or without background knowledge,  the probability that a subscriber can infer locations of users registered with it is nearly 0. 
\end{theo}

{\em Proof:} The subscriber stores the mapping between the real identities and the pseudo  IDs of users registered with it. As the location reporting is done directly between the user's mobile app and the servers but not through the subscriber, the subscriber does not have any location related information of the users. Upon receiving the contact tracing results, the subscriber will know who are  in close contact of the diagnosed patient, but still do not know what places they had been to together  since the server does not return any location information in the query result.  

Also, subscribers will not gain any location information by colluding with one another and exchanging their users' information  since they all receive the same query results that contain only pseudo IDs but nothing about locations.  $\blacksquare$

\begin{theo} With or without background knowledge,  the probability that a user can infer other users' IDs or locations more than his/her  background knowledge is nearly 0. 
\end{theo}

{\em Proof:}  In our system, peer users are not sharing information with each other, and hence they do not know each other's pseudo ID or location information. Even if an attacker compromises multiple users'  mobile apps, the attacker will  only know the victims' pseudo IDs and locations, but still nothing about  others.  $\blacksquare$

\vspace{5pt}
Next, we investigate more complex scenarios in which an attacker has compromised multiple parties within the system.

\begin{itemize}
    
\item {\bf One subscriber and one server under attack}: In the event of a simultaneous compromise of one organization and one server by an attacker, the attacker can determine the pseudo IDs that belong to the compromised organization. Based on this information, the attacker may be able to map the group of pseudo IDs of users in that subscriber to the grid cell of the subscriber's location. However, the attacker will still be unable to determine other locations visited by these users due to the secret sharing protocols implemented on the server side. Additionally, the physical relationships between different grid cells are obscured by the hidden timestamps and visiting orders of the places, preventing the attacker from deducing further information.

\item {\bf One user and one server under attack}:  If an attacker gains access to both a user's device and a server, he/she can obtain information about all the locations and grid cells the user has visited, as well as the pseudo IDs of other users who have visited the same grid cell as the compromised user. However, the attacker will not be able to determine the actual locations of the other places visited by these retrieved pseudo IDs, as this information is stored as secret shares. The attacker will not know the real IDs of these users either because the server does not store any real IDs. 

\item {\bf One subscriber and one user under attack}: Without hacking into the server, the information the attacker can obtain from a compromised user and a compromised subscriber is limited to the victim user's location information and the pseudo IDs of the users registered at the same subscriber. This information is restricted to what the compromised parties have access to.

\end{itemize}

To summarize, in order for an attacker to reveal the true locations of all users, the attacker would need to compromise the majority of the servers, subscribers, or users, which would present a significant financial and resource barrier for the attacker.

\section{Experimental Studies} \label{sec:exp}



The main goal of the experiments is to evaluate the efficiency of the proposed privacy-preserving data insertion and contact tracing queries. For this, we compare  our system with two baseline approaches: (i) the system without any privacy protection (denoted as ``NoProtect"), i.e., directly works on plain texts of user data; (ii) the system with privacy protection but without space partitioning trees, denoted as (``NoTree"). All the experiments are conducted on a desktop with Intel Xeon Bronze 3104 1.7GHz CPU, 64GB RAM.

In the experiments, we use datasets derived from the real dataset called GeoLife \cite{GeoLife}, which contains  17,621 trajectories with a total distance of 1,251,654 kilometers over four years. Each trajectory is represented as a sequence of time-stamped locations with longitude and latitude. These trajectories are originally from 182 users over four years. Since the real dataset is relatively small, we generate synthetic datasets that mimic GeoLife trajectories, in order to test the scalability of our approach. We vary the total number of users from 100K to 500K. Each user's trajectory is generated by  picking locations in the real trajectories where people have lingered for at least the minimum infectious duration. Each  trajectory contains maximum 10 locations. We generate one trajectory per user per day for 14 days. This results in maximum 7M locations and 70M locations in the test dataset. Recall that  our system can take locations collected via any  means including but not limited to GPS and indoor positioning systems.  To ensure consistent calculations across different types of locations, they will be converted to a unified coordinate representation for storage and computation purposes. 

Both the insertion and query performance are evaluated using CPU time. We record the average insertion time per user when recording all the users' daily trajectories. When testing queries, we randomly select 100 users as patients to launch the contact tracing queries, and record the average query time.

\subsection{Effect of the Total Number of Trajectories}

In the first round of experiments, we vary the total number of trajectories from 1.4M to 7M which are corresponding to 100K to 500K users trajectories in 14 days.  The infectious distance is set to 2m, and the incubation period is 14 days. In this round, we adopt a space partitioning with small grid cells of approximately 12m by 9m. Specifically, we first partition the space into 173 region cells, and then further partition each region into 176 cells, which result in a 3-layer space partitioning tree. We compare the performance of our system with a similar system that uses plain text opposed to secret shares to store the data and does not protect privacy.

\begin{figure}[!t]
	\centering
	\includegraphics[width=2.8in]{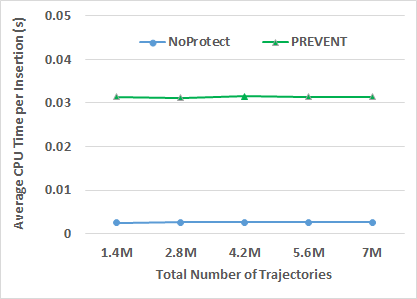}
	\caption{Insertion Time regarding Number of Trajectories}
	\label{fig:traj_insert}
\end{figure}

Figure \ref{fig:traj_insert} reports the average insertion cost of the last 100 users being inserted. It is not surprising to see that our privacy preserving algorithm takes more time than the algorithm that works directly on the plain texts. This is because to achieve privacy preservation, we need to conduct multiple rounds of secure comparison when inserting  user's location information. Fortunately, our algorithm is still fast enough to provide real-time services as each user's  insertion can be completed still within milliseconds. Moreover, our insertion cost stays nearly constant with the increase of the number of users and trajectories. which demonstrates the scalability of our system. It is attributed to the use of our proposed space partitioning tree. For each insertion, we only need to compare the new user with the single representative user in each grid cell. As long as the space partitioning is the same , i.e., the total number of grid cells stays the same, the insertion performance will not be affected by the total number of trajectories that need to be stored.

\begin{figure}[!t]
	\centering
	\includegraphics[width=2.8in]{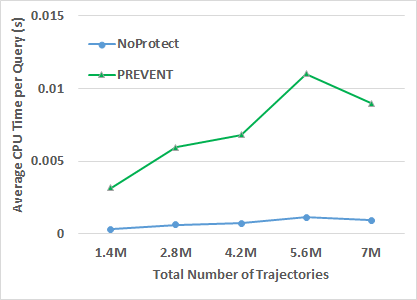}
	\caption{Query Time regarding Number of Trajectories}
	\label{fig:traj_query}
\end{figure}

Next, we examine the query performance of our algorithm against the baseline approach that has no privacy protection. Figure \ref{fig:traj_query} shows the average query cost of finding the people who were within the infectious distance of a given patient during the incubation period. From the figure, we  can observe that the time to perform our secure query is very short, i.e., only a few milliseconds, even thought it is slower than the approach without privacy protection. The overhead of our query algorithm is introduced by  the need to conduct secure comparisons between secret shares of users' trajectories. In addition, we also observe that the query cost of both approaches increase with the total number trajectories. The reason is straightforward. In the same space, the more trajectories, the more people may be within the infectious distance of the patient, and hence more comparisons  are needed.

\subsection{Effect of Space Partitioning }

We now take a closer look at the effect of space partitioning by comparing the performance of our approach using small and large grid cells, respectively. Here we use the dataset with 100K users and 1.4M trajectories. The partitioning with small cells are the same as that in the previous experiments whereby each cell is about 12m by 9m. The partitioning with large grid cells has the cell size of 120m by 90m. Both have three layers. When the small grid cell is used, the entire space is  first partitioned into 173 regions and each region is further partitioned into 176 sub-regions. Then, each sub-region contains  around 197 small cells. When the large grid cell is used for partitioning, the whole space is first partitioned into 33 regions, and each region is divided into 39 sub-regions. Finally, each sub-regions contains approximately 45 grid cells.


	
\begin{figure}[!b]
		\centering
		\includegraphics[width=2.8in]{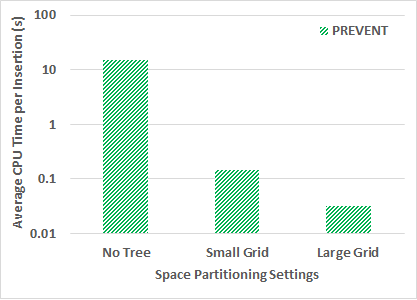}
		\caption{Insertion Time in Different Space Partitions}
		\label{fig:space_insert}
	\end{figure}

	Figure \ref{fig:space_insert} compares  the average insertion cost in the following three scenarios: using no space partitioning tree but only one level of large grid cells, partitioning using small grid cells,  and partitioning using large grid cells.   
	Observe that the insertion cost is highest among all when no space partitioning tree is used, and the insertion cost is lowest when larger grid cells are used. This is because without space partitioning tree, an insertion needs to be compared with the representative user in each grid cell. With the aid of space partitioning tree, an insertion only needs to compare with one grid cell at each level of the space partitioning tree, which significantly reduce the insertion cost. In addition, the larger grid cell also helps reduce the insertion cost. Recall that a user's location near the border of a cell will be included in the neighboring cell as shown in Figure \ref{fig:border}. When the size of a grid cell is large, there are fewer such borderline cases, and hence fewer insertions. 
	

	
\begin{figure}[!t]
		\centering
		\includegraphics[width=2.8in]{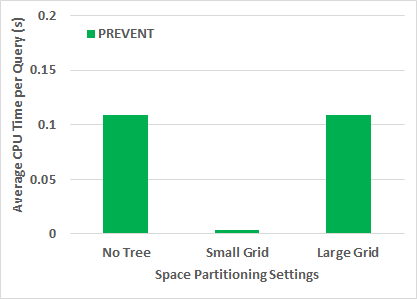}
		\caption{Query Time in Different Space Partitions}
		\label{fig:space_query}
	\end{figure}

	Figure \ref{fig:space_query} shows the corresponding query performance under the same three settings. The query cost is measured using the average CPU time of 100 queries. Observe that the query performance is better when the grid cell size is smaller. This is because the query process compares the patient's trajectory with the trajectories in the grid cells that the patient is located. The larger the grid cells, the more candidate trajectories to be compared, and hence results in longer query time. Also, since the query process uses only the hash table but not the space partitioning tree as shown in Figure 5, the query performance of the NoTree version that based on large grid cells is the same as that of our approach using large grid cells.

	\subsection{Effect of Infectious Distance}
	
\begin{figure}[!ht]
		\centering
		\includegraphics[width=2.8in]{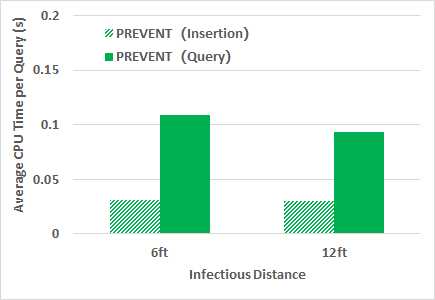}
		\caption{Effect of Infectious Distance}
		\label{fig:infectious}
	\end{figure}

	This round of experiments evaluates the effect of infectious distance which varies from the typical  2m distance to a longer distance of 4m. The dataset used for testing is still the one with 100K users and total 1.4M trajectories. Figure \ref{fig:infectious} reports the insertion and query cost when large grid cells are used for partitioning. 
	The first observation is that the infectious distance does not affect the insertion cost much. This is because the grid cell size is the same. The minor differences in the performance are caused by those data points at the border of cells. As for queries,  there are not significant differences either. This is because the query cost is also determined mainly by the grid cell size. All the users in the grid cell that the patient  has visited will need to be securely compared with the patient's trajectory regardless the length of the infectious distance.

	%
	%
	%
	
	\subsection{Effect of Incubation Period}
	
\begin{figure}[!t]
		\centering
		\includegraphics[width=2.8in]{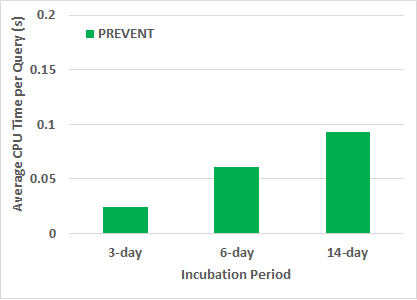}
		\caption{Effect of Incubation Period}
		\label{fig:incubation}
	\end{figure}

	Finally, we study the effect of the incubation period by varying it from 3 days to 14 days. The incubation period only affects the query performance but  not the insertion performance since the space partitioning tree structure stays the same. Figure \ref{fig:incubation} shows the average query cost on the 100K user dataset when using the large grid cell partitioning and 4m infectious distance. As expected, the longer the incubation period, the higher the query cost. This is because longer incubation period requires the query to compare with trajectories across more days, and hence takes more time.
	
	%
	%
	%
	%
	%
	%
	
	\section{Conclusion} \label{sec:con}

	This paper presents PREVENT, a scalable and practical privacy-preserving system for infectious disease contact tracing across multiple organizations. PREVENT effectively prevents any individual party from obtaining precise location information during the entire tracing process, including location collection and location queries. Notably, the system leverages a novel hierarchical query algorithm that delivers real-time performance while ensuring privacy protection. Experimental results showcase its superiority over basic privacy-preserving approaches that lack the sophisticated data structure support present in the PREVENT system. Furthermore, our system is fully extensible, capable of handling hundreds of millions of location data points. 


\section{Acknowledgement}
This work is partially supported by NSF award DGE-1946619.

\bibliographystyle{IEEEtran}
\bibliography{ref}

\begin{IEEEbiography}[{\includegraphics[width=1in,height=1.25in,clip,]{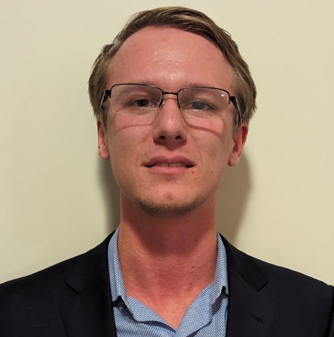}}]{Tyler Nicewarner}
 is currently a PhD Candidate within the I-Privacy Lab at Vanderbilt University. He received a Bachelors of Science degree in Computer Science from the University of Missouri in 2020. His research interests cover many areas in the fields of secure multiparty computation, location privacy, and machine learning.
\end{IEEEbiography}

\begin{IEEEbiography}[{\includegraphics[width=1in,height=1.25in,clip,keepaspectratio]{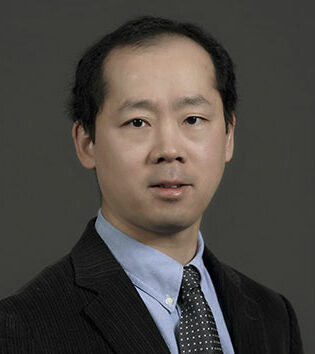}}]{Wei Jiang}
  is a research scientist at Oracle Labs. He obtained his PhD in Computer Science at Purdue University in 2008 and his B.S. in Computer Science and B.S. in Math at University of Iowa in 2002. Before joining Oracle Labs, he was a tenured associate professor at Computer Science Department at University of Missouri. His research interests are applied cryptography and secure multi-party computation in various application domains.
\end{IEEEbiography}

\begin{IEEEbiography}[{\includegraphics[width=1in,height=1.25in,clip,keepaspectratio]{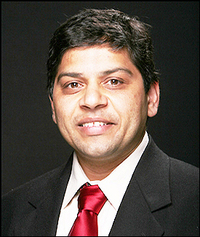}}]{Aniruddha Gokhale} (Senior Member, IEEE) received the B.E. degree from the University of Pune, India, the M.S. degree from Arizona State University, and the Ph.D. degree in computer science from the Washington University, USA. He is a Professor with the Vanderbilt University. His research focuses primarily on solving systems-level challenges by designing and implementing innovative algorithmic solutions incorporating elegant software engineering principles, such as design patterns, domain-specific modeling and generative programming. Specifically, he is interested in solving systems problems involving a variety of quality of service and data consistency issues through effective resource management, particularly in cloud computing, cyber physical systems, and Internet of Things.
\end{IEEEbiography}

\begin{IEEEbiography}[{\includegraphics[width=1in,height=1.25in,clip,keepaspectratio]{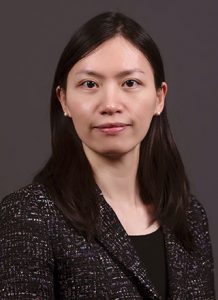}}]{Dan Lin}
 is currently a Professor of Computer Science and Director of the I-Privacy Lab at Vanderbilt University. She received a Ph.D. degree in Computer Science from the National University of Singapore in 2007 and was a postdoctoral research associate at Purdue University for two years. Her research interests cover many areas in the fields of information security and database systems.
\end{IEEEbiography}

\end{document}